\documentstyle[pre,aps,epsfig,preprint]{revtex}
\tightenlines
\def\PRL{Phys. Rev. Lett.}

\begin{document}	
\title{Evolution of avalanche conducting states in electrorheological liquids}
\author{A. Bezryadin, R. M. Westervelt, and M. Tinkham}
\address{Department of Physics and Division of Engineering and Applied
Sciences,\\Harvard University, Cambridge, Massachusetts 02138}
\date{October 21, 1998}
\maketitle
\begin{abstract}

Charge transport in electrorheological fluids is studied experimentally 
under strongly nonequlibrium conditions. By injecting an electrical 
current into a suspension of conducting nanoparticles we are able to initiate 
a process of self-organization which leads, in certain cases, 
to formation of a stable pattern which consists of continuous 
conducting chains of particles. The evolution of the dissipative 
state in such system is a complex process. It starts as an avalanche 
process characterized by nucleation, growth, and thermal destruction 
of such dissipative elements as continuous conducting chains of 
particles as well as electroconvective vortices. A power-law 
distribution of avalanche sizes and durations, observed at this stage 
of the evolution, indicates that the system is in a self-organized 
critical state. A sharp transition into an avalanche-free state 
with a stable pattern of conducting chains is observed when the 
power dissipated in the fluid reaches its maximum. We propose 
a simple evolution model which obeys the maximum power condition 
and also shows a power-law distribution of the avalanche sizes.
\end{abstract}

\vspace{0.5cm}

\section{Introduction}

A breaking of translational or temporal symmetry often occurs if a homogeneous,
spatially extended system is driven far from equilibrium. This results in 
pattern formation~\cite{Cross}. The patterns (sometimes called ``dissipative 
structures'') accelerate the energy dissipation and the motion of the system 
towards equilibrium. Spatiotemporal disorder which occurs if the patterns 
vary in time and space can involve the chaotic evolution of an amplitude 
field~\cite{Steinberg,Kolodner}, or it can be connected with the dynamics of 
defects~\cite{Porta}. Also, out-of-equilibrium driven systems with 
{\it threshold} dynamics exhibit a rich phenomenology, from synchronized 
behavior~\cite{Strogatz} to self-organized criticality (SOC)~\cite{Bak,Olami}, 
when the long-range correlations are manifested as power-law distributions of 
avalanche sizes and lifetimes~\cite{Held,Westervelt}.

In this paper we study the evolution of dissipative structures in initially
homogeneous electrorheological fluids~\cite{Havelka} suddenly driven out of 
equilibrium by applying a strong electric field. The driving 
mechanism of the evolution is found to be a competition between the 
forces which attempt to order the system and the destructive influence of 
increased thermal fluctuations. The ordering forces appear when the system
is driven out of equilibrium. In our case it is the electric field which 
polarizes the particles and leads to dipole-dipole attraction between 
them. The ordering leads to an increase in the dissipation rate.
The increasing rate of the dissipation and associated temperature rise
have an opposite, destructive effect on the self-organized structures. 

Our attention will be restricted only to systems with {\em limited} 
dissipation. They consist of two parts: an ``adaptive'' subsystem 
(the electrorheological fluid) and a ``rigid'' subsystem (in our experiments 
it is a plain resistor connected in series with the fluid). The rigid part 
imposes an absolute limit on the power dissipated in the fluid. As a 
consequence, there is a global nonlinear interaction between all dissipative 
elements of the forming dissipative structure. Two types of collective behavior 
which lead to an increase in the dissipation rate have been encountered. 
These are (i) the conducting chains~\cite{Halsey} which appear due to the 
dipole-dipole attraction and (ii) convective flows of electrically 
charged volumes of the liquid (see the illustration in Fig.1). 
The degree of order is characterized by the electrical current, which we 
can accurately measure. The charging of nanoparticles and the associated 
{\em repulsion} between them competes with the dipole-dipole
{\em attraction} and renders the chain formation less evident than in
usual electrorheological liquids with zero conductivity~\cite{Martin} where
the charging is not possible. 

New findings/results presented in this article are the following: 
(i) Two qualitatively different current-carrying states are found in 
electrorheological liquids exposed to a strong electric field. 
A ``scale-invariant'' avalanche state (AS) appears at the beginning 
of the evolution. It resembles the SOC state observed previously in, 
for example, sandpiles~\cite{Held} and is characterized by a power-law 
distribution of avalanche sizes and durations (even though there is no external 
flux-drive~\cite{Sornette}). (ii) The AS can transform itself into a 
stable state (SS) with a visible pattern of strings of nanoparticles 
(Fig.2). (iii) This transformation (which can be considered as a pattern 
formation) takes place only if the power dissipated by the adaptive part 
(the fluid) reaches its maximum (imposed by the rigid part). (iv) We propose 
a simple evolution model which obeys the ``maximum power'' 
principle~\cite{Odum,Nagel} and shows an avalanche state with a power-law 
distribution of sizes and durations.

\section{Experimental details}

The sample configuration is depicted in Fig.1. It consists 
of a pair of stainless-steel parallel cylindrical electrodes, 
0.7 mm in diameter, separated by a distance of 10 mm and 
immersed over 10 mm into an electrorheological fluid. The fluid 
consists of a dielectric solvent (toluene) with ultrasonically 
dispersed conducting carbon nanoparticles~\cite{Carbon} available 
commercially~\cite{Carbon1}. The concentration of particles 
is $\approx 0.02 mg/ml$, which is far below the percolation threshold. 
Consequently, the initial resistance between the electrodes is high 
($\sim 10^{12}\Omega$). At time $t=0$ a DC voltage $V_{0}=100V$ is 
applied to the electrodes through a series resistor $R_{s}$ (Fig.1). 
Then an evolution curve, current vs time, $I(t)$, is measured. 
In the following discussion we will distinguish between curves 
measured on the ``same sample'', and on ``different samples''. 
In the first case a series of $I(t)$ curves is measured on the 
same hermetically closed bottle with electrodes and the fluid. 
To restore the homogeneity, the fluid is excited ultrasonically 
before each new $I(t)$ measurement~\cite{Expl}. Measurements on 
different samples means that a new, freshly prepared suspension 
is used for each new sample.

\section{Transport measurements}

Experimentally we find three main Evolution Scenarios. 
(i) ES1: The first measurement on a freshly prepared suspension 
shows a monotonic growth of the current with time, if the 
concentration of particles is high enough. An example of such 
behavior is given in Fig.3a, curve ``A''. (ii) ES2: Next 
measurements on the same sample show much more complicated 
curves with three different stages. For example, the curve ``B'' 
in Fig.3 was measured on the same sample as curve ``A'' after 
the fluid was again homogenized ultrasonically. Curves ``C'' and ``D'' 
were taken one after the other on a different sample, 
using a much higher series resistor $R_{s}$. They also 
illustrate the scenario ES2. Three different stages observed in 
the ES2 case are described below. Stage 1: During the first few 
hundreds of seconds (or less) after the voltage is applied, 
the current is small and it does not grow considerably; Stage 2 
(avalanche state, AS): strong fluctuations of the current (by a 
factor of $\approx100$ in some cases) appear; the averaged value 
of the current ($<I>$) gradually increases. Stage 3: The current 
rapidly increases to yet a higher level and the fluctuations 
disappear. Thereafter the current continues to grow very slowly 
and monotonically. 
This new stable state (SS), which is the final stage of the 
evolution, is characterized by a visible and stable pattern of 
entangled strings composed of carbon particles. 
Examples of such strings are visible on the four bottom pictures 
of Fig.2. (iii) ES3: After a few successive measurements on the 
same sample the system can not reach the stable state any more, 
but the first two evolution stages are the same. Examples of such 
evolution curves are shown in Fig.3b, curves ``E'' and ``F''. These 
curves were taken after the curves ``C'' and ``D'', on the same sample. 
The avalanche state in ES3 case lasts $\sim 10^{5}s$ and finally, 
instead of the transition to the stable state, the current slowly 
decreases to zero; the conducting state ``dies''. 
This happens when most of the particles  cluster and settle down. 
(iv) After many ($\sim 10$) measurements, the same sample shows 
no current growth at all.

The described evolution scenarios are quite general. They have 
been observed in liquids with different viscosity (toluene, 
hexadecane, mineral oil), with different electrodes (e.g. Pt, 
Sn), and at different values of the series resistors $R_{s}$. 
We have also found that the evolution scenarios described above 
can be observed not only by doing repetitive $I(t)$ measurements 
on the same sample (this ``aging'' technique was described above) 
but also by reducing the concentration of the nanoparticles, 
while the I(t) curve is measured only once on each new sample 
with a freshly prepared suspension. The concentration reduction 
leads to the same transitions between evolution scenarios as the 
aging (when a series of $I(t)$ measurements is made on the same 
sample). The aging approach was found to give much more 
reproducible results than the approach when the concentration 
is the control parameter.

\section{Imaging of the pattern formation process}

The evolution of patterns in electrorheological liquids can be 
observed directly with an optical microscope. Photographs shown 
in Fig.2 illustrate different stages for the evolution of the type 
ES2. The first image shows an aggregation process which takes place
in the suspension of particles at zero electric field. Formation 
of fractal-like clusters is clearly visible. 

The voltage was applied at time $t=0$ between the two
electrodes (black). The applied field causes a strong 
polarization of the clusters (made of electrically conducting 
particles). The second photograph in Fig.2 shows that at 
$t \approx 2 \ s$, $i.e.$ immediately after the voltage was applied, 
all big aggregates break apart, so the fluid looks much more uniform. 
This rupture process is due to the polarization mentioned above.
Since opposite sides of polarized clusters of nanoparticles carry 
opposite charges, big enough clusters are pulled apart if the 
applied electric field is strong enough. 

During the first few hundreds of seconds the system shows some sort 
of collective behavior which may be called electroconvection 
or a ``shuttling'' effect. At this stage the electrical current 
is carried from one electrode to the other by macroscopic streams 
which develop in the fluid. Each stream carries many charged
particles (or small clusters of particles) with the same charge. 
Upon the contact with the oppositely charged electrode, 
the particles acquire the opposite 
charge and start to move toward the opposite electrode. Initially 
those streams are very unstable and the flow looks ``turbulent''. 
With time, new streams nucleate, become stronger and disappear. 
This shuttling effect is shown schematically in Fig.1. At this stage 
no stable strings were observed.

As time passes, the streams become bigger and slower.
At some moment we observe an abrupt ``stabilization'' transition 
(which takes less than a second) when the turbulent 
electroconvection disappears and continuous strings of particles, 
extended from one electrode to the other, become visible. This is 
illustrated in Fig.2 (third image) which is taken a few seconds 
after the first stable strings became visible. Note that though we do 
not see the strings before the stabilization transition, we can not 
exclude that some strings of particles are being formed for a short 
time and then destroyed by heating or convection. After the pattern 
is stabilized, the strings show a tendency to form bundles. As is 
shown in the forth, fifth, and sixth images of Fig.2, these bundles grow 
continuously with time (which leads to the measured monotonic decrease of 
the sample resistance).

\section {The maximum power principle}

The electrical scheme of our setup is shown in Fig.1. 
The power dissipated by the electrorheological fluid can be 
written as $P_{f}=I(V_{0}-V_{s})=4P_{max}/(r+2+1/r)$ 
where $P_{max} \equiv V_{0}^{2}/4R_{s}$, $r \equiv R_{f}/R_{s}$, 
$R_{f} \equiv V_{f}/I$ is the time-dependent resistance of the fluid, 
$V_{0}$ is the battery voltage applied to the fluid and the 
resistor $R_{s}$ connected in series (see the schematic in 
Fig.1), $V_{s}$ ($V_{f}$) is the voltage drop on the series 
resistor (fluid), and therefore $V_{0}=V_{s}+V_{f}$. 
The expression for the $P_{f}$ has a single maximum which is achieved when 
$r=1$ or $R_{f}=R_{s}$. Therefore the maximum power which can be 
dissipated by the fluid is $P_{max}=V_{0}^{2}/4R_{s}$. 
Note also that the expression for the $P_{f}$ is symmetric 
under substitution $r \rightarrow 1/r$ or, what is the same, 
$R_{f} \rightarrow R_{s}^{2}/R_{f}$. In other words, any allowed 
($P_{f} \leq P_{max}$) level of the dissipated power $P_{f}$ 
(except only one point $P_{f}=P_{max}$) can be achieved in two 
physically different states of the fluid. Our measurements 
show that these two states are qualitatively different. 
All states with $R_{f}>R_{s}$ are characterized by
strong avalanche-like current fluctuations. As soon as the 
fluid resistance decreases to the level $R_{f}=R_{s}$ where 
dissipated power reaches its absolute maximum, the fluctuations
disappear abruptly. At $R_{f}<R_{s}$ the fluid resistance continues
to decrease but slowly and monotonically. In Fig.4 we plot the power 
dissipated by the fluid (normalized by $P_{max}$) versus time. Curves 
``G'' and ``I'' (which correspond to the ES2 scenario) illustrate 
the maximum power principle for two different values of the series 
resistance $R_{s}=48.1M\Omega$ (curve ``G'') and $R_{s}=1.04G\Omega$
(curve ``I''). In both cases the huge current fluctuations
disappear when the power reaches the maximum when $P_{f}/P_{max}=1$. 
The curve ``H'' shows the normalized power vs time in the ES3 case. 
In this evolution scenario the power does not increase up to the 
$P_{f}=P_{max}$ level. Consequently the system never stabilizes. 
To summarize, the experiment shows that the choice between the two 
scenarios (ES2 or ES3) is determined by the ability of the adaptive 
part of the system with limited dissipation to reach the maximum 
rate of energy dissipation.

\section{Avalanche state and self-organized criticality}

It is interesting to compare the dynamics of fluctuations, 
observed in our systems, to the critical behavior of sandpiles 
and other self-organized systems. We suggest that the huge current 
fluctuations measured before the pattern is stabilized, constitute 
an avalanche activity of the dissipative structure. To make a 
quantitative comparison, we analyze the distribution 
of avalanche sizes ($X$) defined 
as the amplitude (in Amperes) of each monotonic decrease of the current. 
This definition is acceptable since the noise level of our apparatus 
($<1 \ pA$) is much lower than the amplitude of the current fluctuations.
Similarly, the duration of an avalanche $T$ is defined as the duration 
(measured in seconds) of each monotonic current drop.

Statistical analysis shows that the avalanche activity in our 
system is scale invariant. This means that the avalanche distributions 
do not peak at any particular value. In the example of Fig.5b 
(see triangles), the avalanche-size probability density $D_{X}$ follows 
a power-law distribution $D_{X}\sim X^{-\alpha}$ (with $\alpha\approx 1$) 
over about four decades. This suggests that the dissipative structure 
(before it is stabilized) is in the self-organized critical state. 
To corroborate this, we have found the distribution $D_{T}$ of avalanche 
durations which is plotted in Fig.5c. It is also a power-law 
distribution:  $D_{T}\sim T^{-\beta}$, as can be expected for a 
self-organized critical state. The exponent is larger in this case:
($\beta \approx 2.3$). Other samples have shown a very similar behavior. 

The distributions of avalanche sizes and durations, presented above, 
have been calculated for the evolution curves of the type ES3.
In this cases the avalanche state lasts up to $10^{5} \ s$.
In the ES2 case, the avalanche state lasts for a much shorter period 
of time, but the distributions are similar to those in the ES3 case.

\section{Dissipative elements as building blocks of the dissipative 
structure}

Many properties of the pattern evolution, described above, can be 
understood by introducing the notion of ``dissipative elements'' (DE).
The dissipative structure is assumed to be composed of relatively 
independent dissipative elements. In general, a DE is a region of 
space where any sort of self-organization or collective behavior
(in our system it may be chain formation or electroconvection) leads 
to a strong increase of the local dissipation. The capability of each
DE to dissipate energy $G_{i}$ (which is electrical conductance in our
case) as well as the total number of DE's are assumed to grow with time. 
This reflects the general tendency of ordering, observed in 
nonequilibrium systems. This tendency will force more and more 
particles to join the existing dissipative elements or to form
new ones. This process of ordering may be limited by the heating 
associated with the activity of each DE. To build a simple evolution 
model (see below) we will assume that any DE burns out when the 
power dissipated in it reaches some critical value $P_{c}$. 
This constitutes the threshold dynamics of our system.  

Under the assumptions, outlined above, the evolution consists of 
nucleation, growth, and destruction of dissipative elements. 
The stabilization transition, observed experimentally, can be 
understood in the following way. In the case when the total rate 
of the dissipation is limited (by the presence of a series resistor 
in our case), the pattern stabilizes if a sufficiently big number 
of DE's with high enough $G_{i}$ values develops at the same time.
In this case the total dissipated power (which is never bigger than
$P_{max}$) will be shared between a big number of DE's. Therefore 
the dissipation in each DE ($P_{i}$) can never become strong
enough for it to burn out.  
 
It has to be explained why the stabilization coincides with the 
point of maximum power and does not depend on the properties of the 
fluid. This follows from the fact that
$P_{f}=4P_{max}/(r+2+1/r)$ and therefore the decreasing resistance
(or increasing conductance) of the fluid causes an increase 
in the dissipation rate {\it only} if $r>1$ or $R_{f}>R_{s}$. 
Oppositely, if $R_{f}<R_{s}$ then the increasing degree of order 
and associated decrease in the fluid resistance lead 
to a {\it decrease} in the power dissipated in the fluid. 
Therefore the pattern stabilizes as soon as the power reaches the 
maximum. We assume here that
the degree of order always increases with time if the system is 
driven far enough from the equilibrium (meaning in our cases that 
the applied voltage is strong enough). Also we assume that the 
ordered structures may be destroyed due to the heating,
but {\it only} if the local power reaches some critical value 
(as was already explained above).

\section{Evolution Model}

To confirm our hypothesis that the experimentally observed behavior
is caused by nucleation, growth and destruction of DEÕs by the 
local heating associated with each DE, we suggest the following 
simple evolution model formulated in terms of electrical circuits. 
Let $R_{i}$ to be electrical resistance of the $i^{th}$ dissipative 
element. The conductance $G_{i}=1/R_{i}$  represents the efficiency 
of the $i^{th}$ DE to dissipate energy. The number of particles 
joining each DE increases with time and so $G_{i}$ increases as well. 
All DEÕs are assumed to be connected in parallel. We consider a 
model where the power dissipated by all DEÕs together can not exceed 
some value $P_{max}$. In the model (as well as in the experiment) 
the power is limited by a resistor $R_{s}$ connected in series with 
DEÕs. The total current can be written as $I=V_{0}/(R_{s}+1/G_{f})$ 
where the total conductance of the fluid is  
$G_{f}=\sum_{i=1}^{N} G_{i}$. The sum is taken over all available 
dissipative elements. Their total number will be $N$ ($N>>1$), 
but some of them may be switched off (meaning that $G_{i}=0$ for them). 
The ``threshold dynamics'' appears due to the assumption that 
the heating destroys the order in the $i^{th}$ dissipative element 
and its conductance goes to zero if the power $P_{i}=V_{f}^{2}G_{i}$ 
dissipated by this particular DE exceeds some critical value 
$P_{c}$ ($P_{c}<<P_{max}$). Since all DE's are assumed to be connected 
in parallel, they all will be biased with the same voltage 
$V_{f}=V_{0}-IR_{s}$ which depends on the total conductance of the 
fluid ($G_{f}=1/R_{f}$). Clearly this causes a global (and nonlinear) 
interaction between all DEs. Indeed, if one of the DEs burns out, 
then $V_{f}$ increases and therefore some other DEÕs with 
a high conductance may burn out as well. 
This leads to further increase of the voltage $V_{f}$ applied to the
electrodes and may lead to destruction of other DE's. 
Such a ``chain reaction'' can explain the avalanches observed 
experimentally. 

Our numerical model works as follows. At t=0 all dissipative elements 
have zero conductance ($G_{i}=0$). Each time step we choose randomly 
$N_{1}$ integer numbers $K_{m}$ such that $1\leq K_{m}\leq N$. Some of 
$K_{m}$ numbers may be identical. Here $N_{1}$ is a fixed number, 
such  that $1\leq N_{1} \leq N$. It controls the nucleation rate of 
dissipative elements. 
The $K_{m}$ numbers represent dissipative elements the conductance of 
which is going to be increased during the time step. The conductance 
of DEÕs with corresponding numbers $K_{m}$ is increased in the 
following way: $G_{K_{m}}$ $\rightarrow$ $G_{K_{m}}+RND$. 
Here $RND$ is a random value, such that 
$0<RND<STEP$, and $STEP$ is a constant representing the growth 
rate of DEÕs. Therefore the growth of each DE is a ``biased random walk``.
If two numbers $K_{m}$ are equal then $G_{K_{m}}$ will be increased twice, 
and so on. After the conductance of all randomly chosen DE's is 
increased, following the algorithm explained above, we calculate the power 
dissipated in each DE using the expression $P_{i}=V_{f}^{2}G_{i}$.
If a dissipative element for which $P_{i}>P_{c}$ is found, its conductance 
is put to zero, representing the destruction of this DE. After each such
destruction event the voltage $V_{f}$, which is the same for all DE's, 
is updated. We proceed to the next time step only when there are no 
DE's with $P_{i}>P_{c}$ left. 

It is remarkable that this simple model can produce evolution curves 
which are very similar to the experimental ones. Three examples of the 
power versus time dependence are shown in Fig.6. In all these examples
the DE nucleation rate $N_{1}=5$ and the critical power 
$P_{c}=1.8P_{max}/N$ are the same. The parameter which is changed is the
DE growth rate $STEP$. If the growth rate is low enough, the model
generates a smooth evolution curve without avalanches, which look similar
to the experimental curves of the type $ES1$. Such an example is given
in Fig.6a. The absence of the avalanche activity is due to the low rate
of the conductance growth, which means that a large number of DE's can form
before any particular DE reaches its critical power point. So the total 
dissipated power can reach the maximum before any DE burns out.
After the total power reaches its maximum, no dissipative elements 
will be destroyed because the probability that the rate of the heat 
dissipation in each particular DE would increase goes to zero. 

At higher growth rates (see Fig.6b) the model generates more complicated 
evolution curves. Now it shows the transition from an avalanche to the
stable state, similar to the experimentally measured ES2 scenario. 
If the growth rate is chosen to be yet higher (Fig.6c), the normalized 
power ($P_{f}/P_{max}$) always stays well below unity. Since it 
never reaches the maximum ($P_{f}/P_{max}=1$), the stabilization can 
not be achieved. Consequently the curve shown in Fig.6c represents 
the ES3 scenario when the avalanche state lasts indefinitely long. 

In the framework of our model, the same three types of behavior 
(ES1, ES2, and ES3) can be observed if the growth rate is kept constant
while other parameters are changed. For example, at low values of the 
critical power we always observe the ES3 scenario. By increasing the 
normalized critical power $NP_{c}/P_{max}$ it is possible to shift the 
system to the ES2 and even to the ES1 scenario at yet higher values of the 
normalized critical power.
The maximum power principle is always obeyed: The stabilization takes 
place only if the power can reach the absolute maximum. The model-generated 
evolution curves are also characterized by a power-law 
distribution of avalanche sizes (see Fig.5b, solid dots). The power-law 
exponent $\alpha_{model}\approx 2$ is higher than the experimental value 
($\alpha\approx 1$). In many systems described previously
by other authors an opposite relation was observed when the 
theoretically predicted value for the exponent $\alpha$ was smaller 
than the values observed in experiments.

The model suggests three main parameters which control the transitions
between different evolution scenarios. These parameters are the rates 
of nucleation and growth of dissipative elements and the normalized critical 
power $NP_{c}/P_{max}$ of the DE's destruction. Experimentally we observe 
different evolution scenarios by aging the sample (see the discussion above).
It is not well established which one of the control parameters changes 
during the aging. Preliminary observations suggest that in the 
process of the chain formation the particles can form stable clusters 
which can not be dissociated during subsequent ultrasonic excitation. 
This irreversible clustering leads to a decrease of the total number of 
independent particles participating in the chain formation, and 
consequently causes an effective decrease of the parameter $N$ which 
represents the maximum number of chains. The normalized critical power 
$NP_{c}/P_{max}$ decreases with decreasing $N$. This is one possible 
explanation for the aging process described in Section III which 
leads to the observed transitions from ES1 to ES2 and subsequently to 
the ES3 type scenario.

Our model possesses certain similarities to the models developed by 
D. Sornette~\cite{Sornette}. He proposed a class of models in which 
the self-organized nature of the criticality stems from the fact 
that the critical point (defined as the point when the coherence 
length becomes infinite: $\xi \rightarrow +\infty$) is attracting 
the nonlinear feedback dynamics. His models are based on the 
existence of a feedback of the order parameter on the control 
parameter. Our model also possesses certain feedback mechanisms since 
the order parameter, say dissipated power, tends to destroy
the order in the system. This leads to a decreases of the conductance
and therefore causes a change of the voltage applied to the fluid,
which can be considered as a control parameter.
On the other hand our model is different since it is not 
spatially extended in the usual sense. In our model each dissipative 
element interacts with {\it all} other DE's with the same 
strength, not only with neighbor DE's. Therefore our model 
may be considered as a zero-dimensional one, so that the notion 
of critical state (which is used in Sornette's models) 
defined as the state when the coherence length
diverges ($\xi \rightarrow +\infty$), is not applicable to our 
system. Our model is based on an assumption that the order which 
develops in some {\it nonequilibrium} system may cause 
its own destruction due to the heat dissipated by the ordered 
structures themselves.

\section{Conclusions}

In conclusion, we present an experimental study of the evolution of 
patterns in a system with limited dissipation. Experiments are done 
on a new type of electrorheological fluid. A transition from the SOC-type 
scale-invariant avalanche state to a stable pattern is observed. It takes
place when the  power dissipated in the adaptive part of the system 
reaches its maximum defined by the rigid part. A general model of the pattern 
evolution in nonequilibrium systems with limited dissipation is suggested.

We thank D. Weitz and S. Maslov for useful discussions. 
This work was supported in part by NSF Grants DMR-94-00396, 
DMR-97-01487, and PHY-98-71810.

\pagebreak 
\begin{figure}[h]
\vspace{2cm}
\centerline{ \epsfxsize=7cm \epsfbox{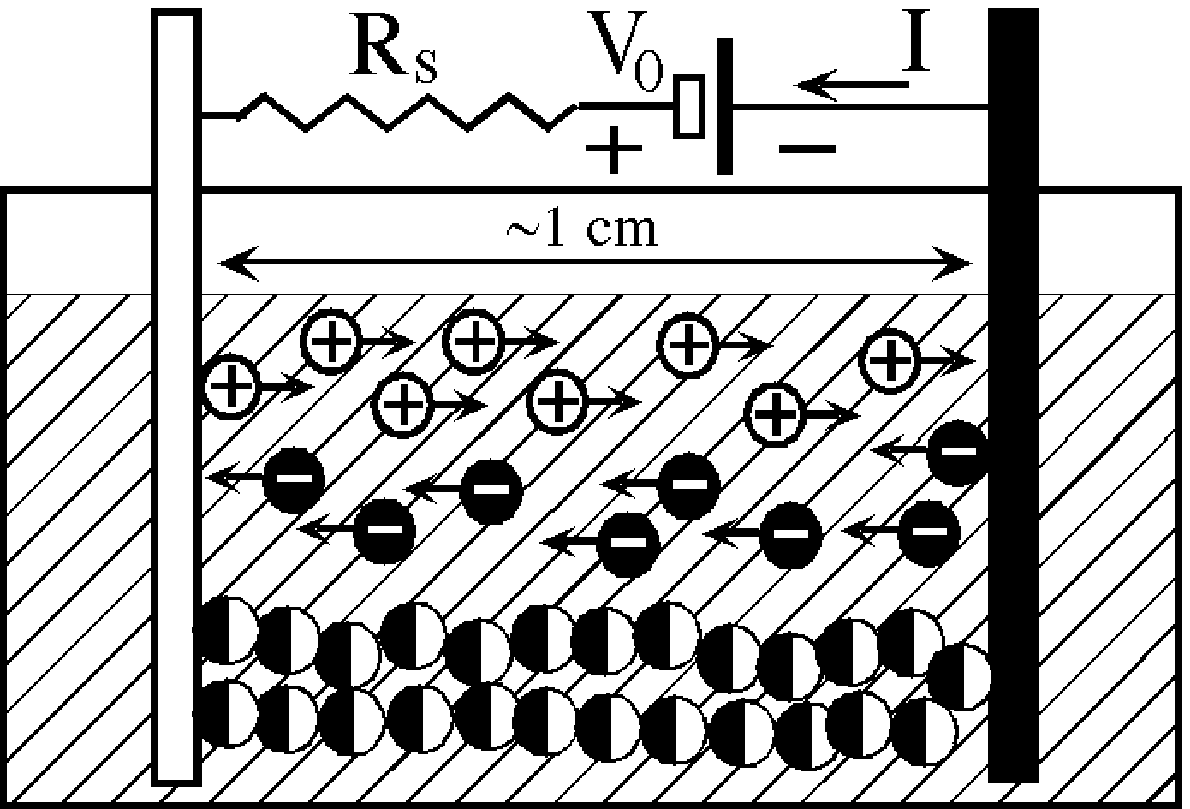}}
\end{figure}
\parbox[t]{16cm}{\small FIG.1 \ \ Schematic of the experimental setup. 
Two electrodes are immersed into electrorheological liquid (dashed region) 
and biased with $V_{0}=100V$. The charging of particles at the electrodes 
causes the electroconvection. Also, polarization of particles by the 
applied electric field leads to chain formation. 
The dissipated power is limited by the series resistor $R_{s}$.} 

\pagebreak
\begin{figure}[h]
\vspace{2cm}
\centerline{ \epsfxsize=7cm \epsfbox{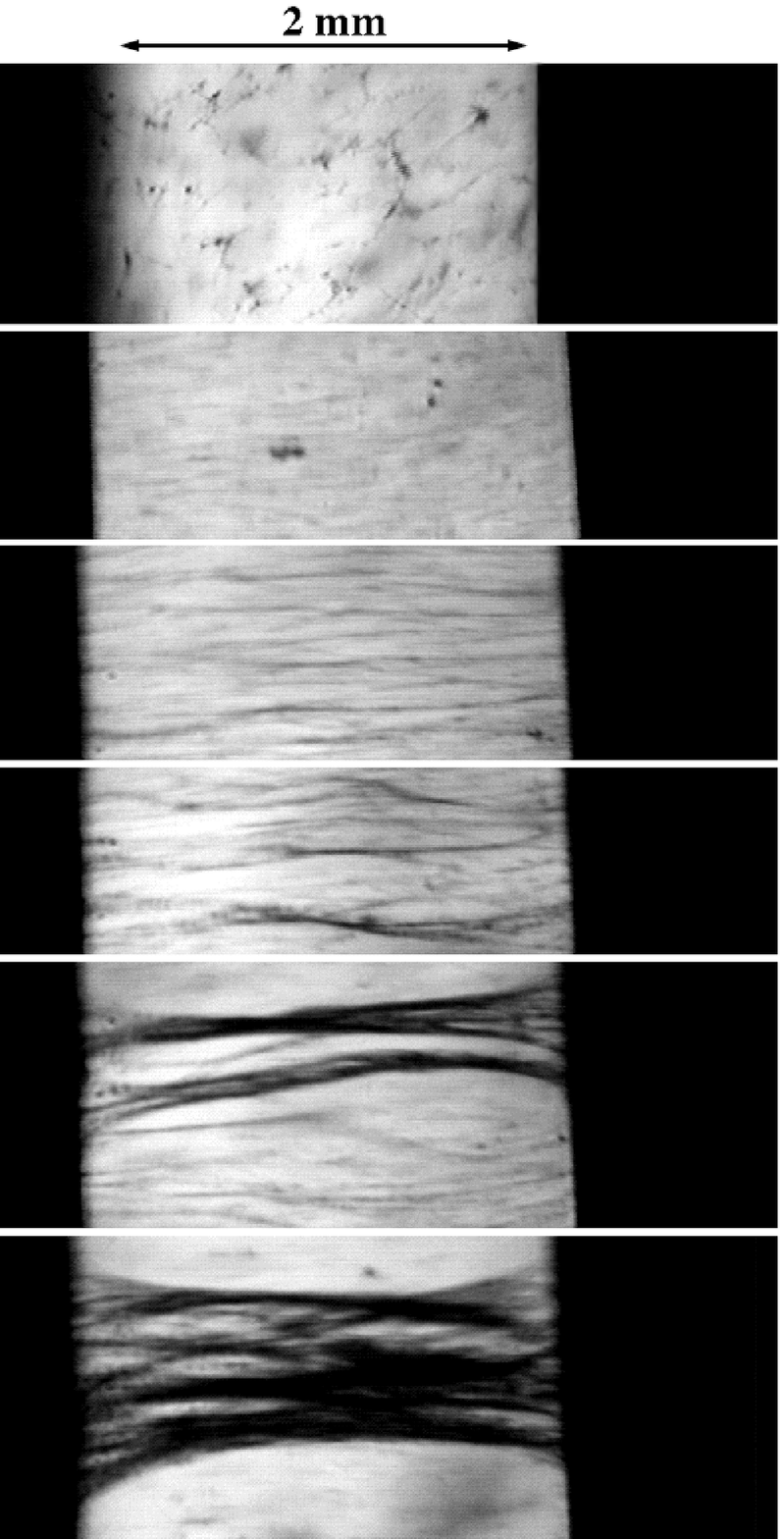}}
\end{figure}
\parbox[t]{16cm}{\small FIG.2 \ \ A time series of sample photographs, 
which illustrate the process of pattern formation. The electrodes are 
visible as black regions on the left and right sides of each image.
The spacing between the electrodes is $\approx 2 mm$. The first 
picture is taken before the voltage was applied ($V = 0 \ V$, $t<0$). 
The following images show the process of pattern formation between 
the biased electrodes ($V=100\ V$). They are take at $t= 2, 300, 310, 
350$, and $400$ seconds respectively. The third picture ($t=300 \ s$) 
is taken a few seconds after visible chains appeared for the first 
time. Subsequent photographs show the tendency of 
bundling. \vspace{0.5cm}} 

\pagebreak
\begin{figure}[h]
\centerline{ \epsfxsize=8cm \epsfbox{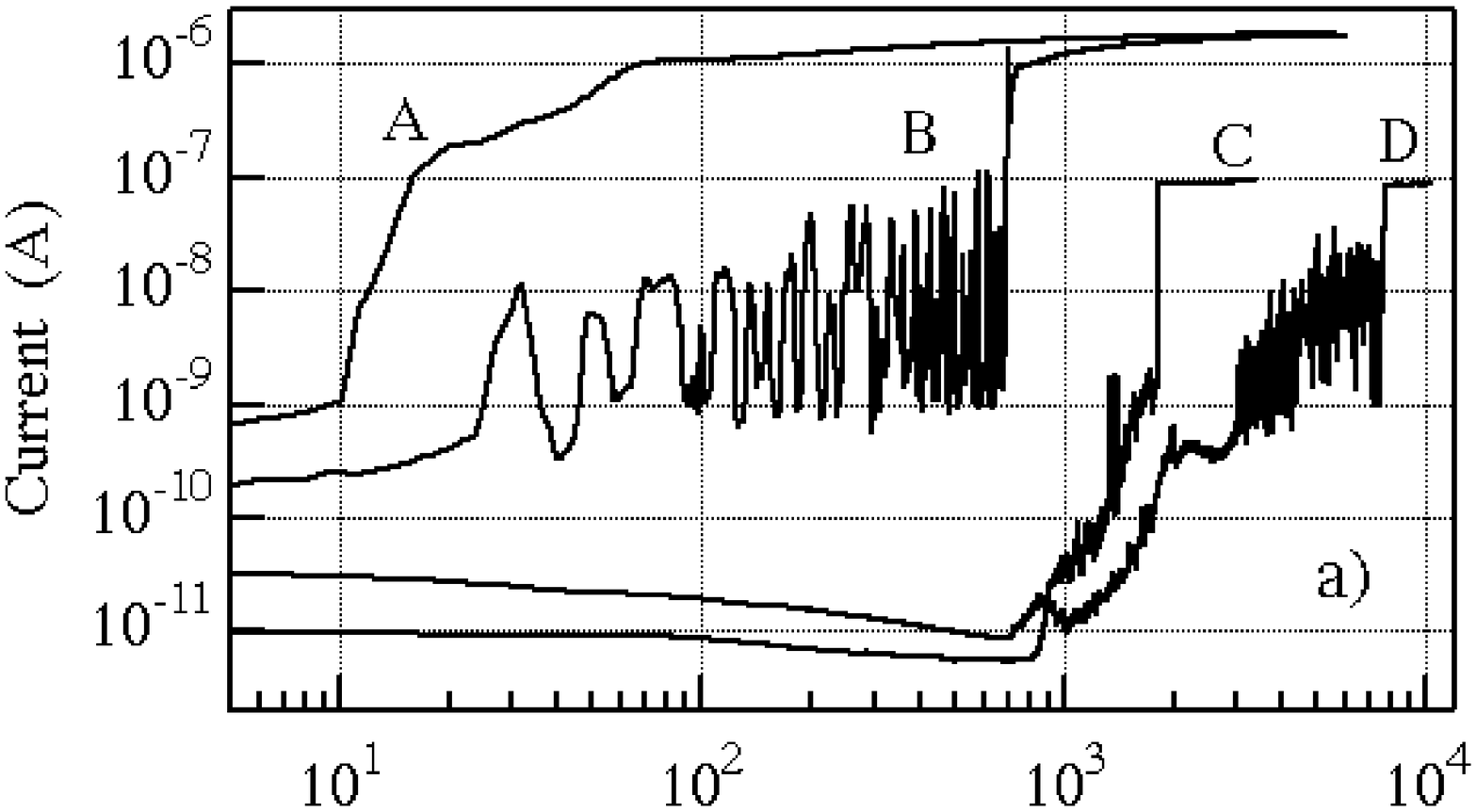}}
\centerline{ \epsfxsize=8cm \epsfbox{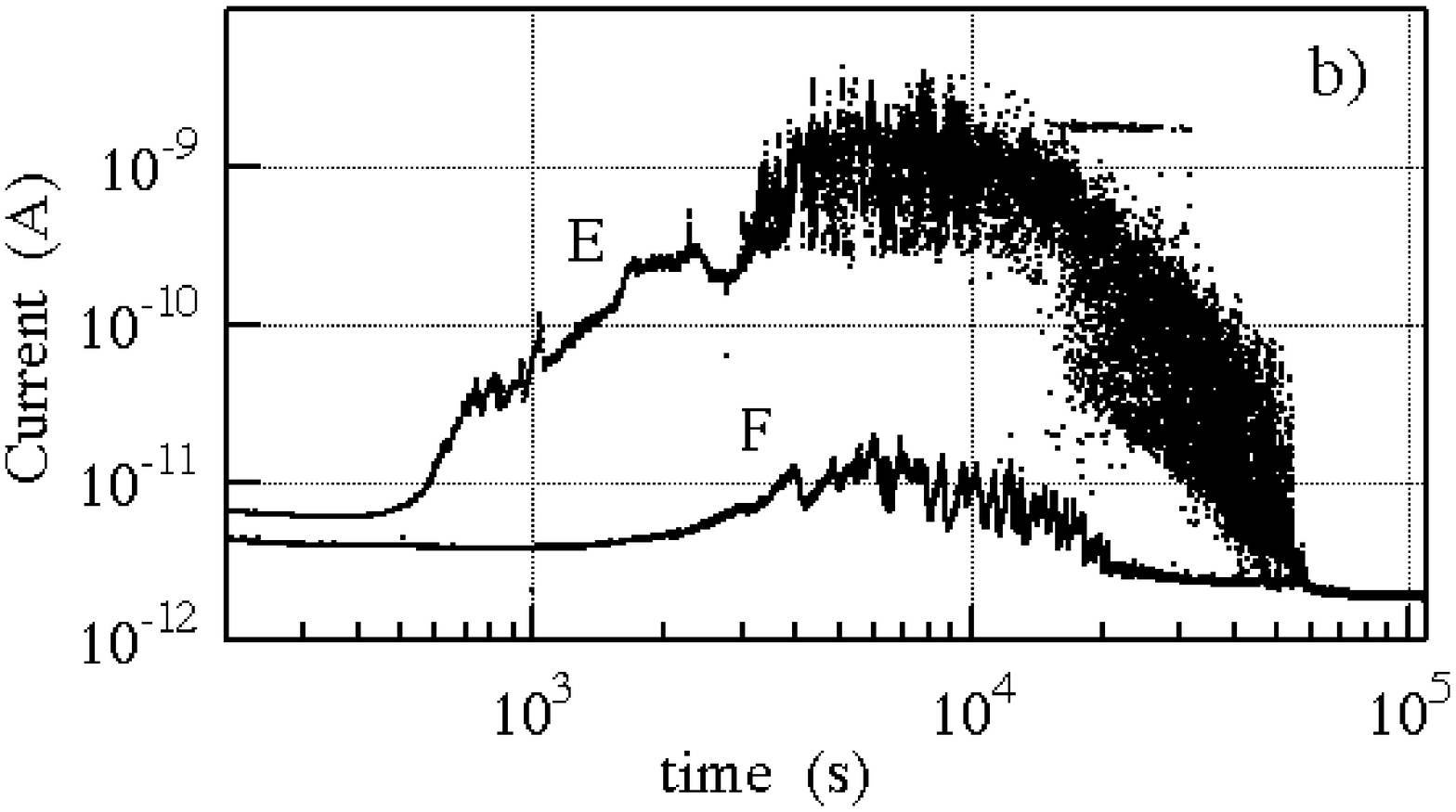}}
\end{figure}
\vspace{-0.4cm}
\parbox[t]{16cm}{\small FIG.3 \ \ Examples of the evolution curves 
(current vs time) for two samples. (a) The curve B is measured after 
the curve A on the same sample with $R_{s}=48.1 M\Omega$ and $V_{0}=100V$. 
Curves C, D, E, and F show successive runs on another sample with 
$R_{s}=1.04 G\Omega$ and $V_{0}=100V$ (curves E and F are plotted in 
the part (b)). The particle concentration was $\approx 0.02 mg/ml$
in all cases.}   

\pagebreak
\vspace{0.3cm}
\begin{figure}[h]
\centerline{ \epsfxsize=8cm \epsfbox{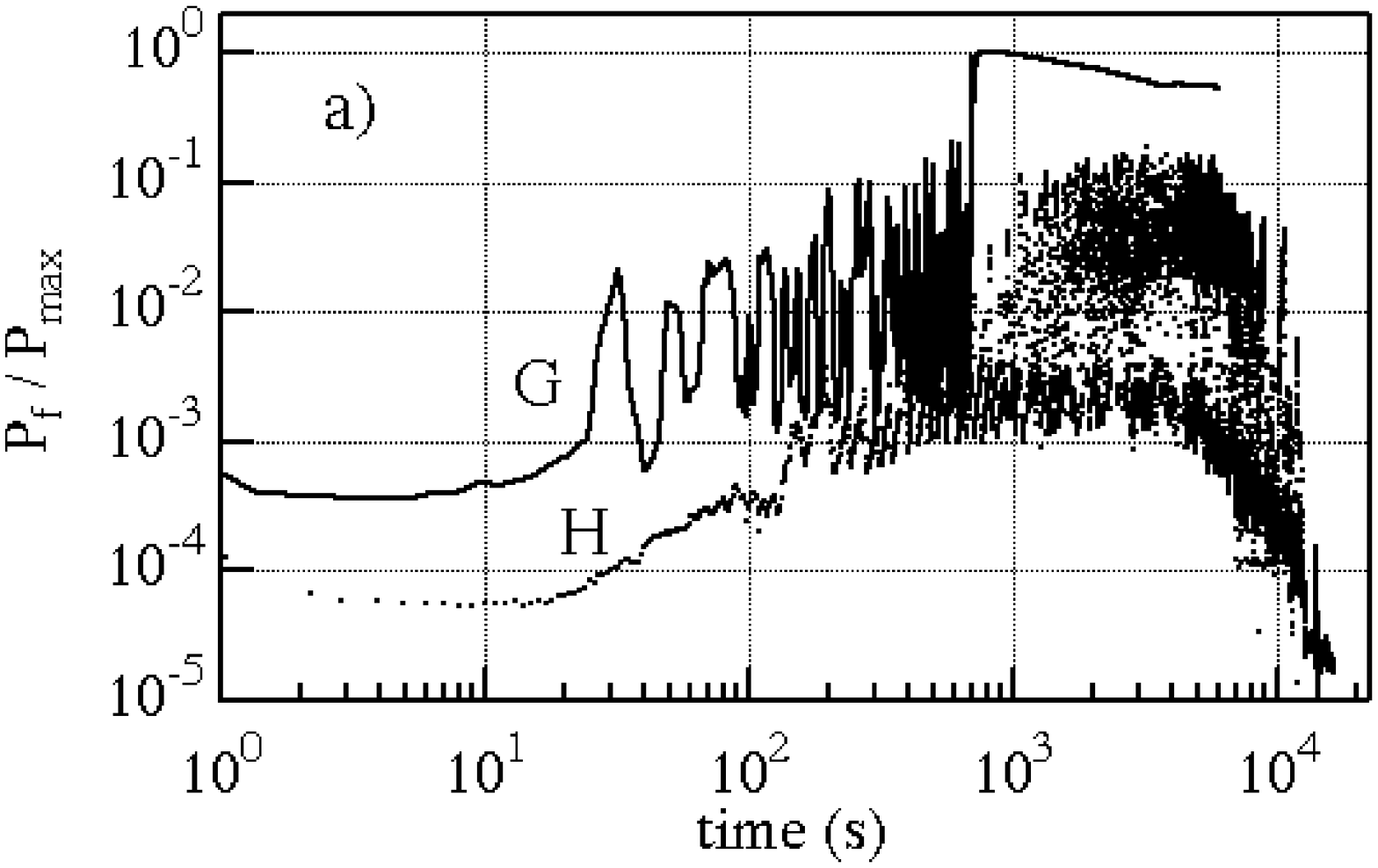}}
\centerline{ \epsfxsize=8cm \epsfbox{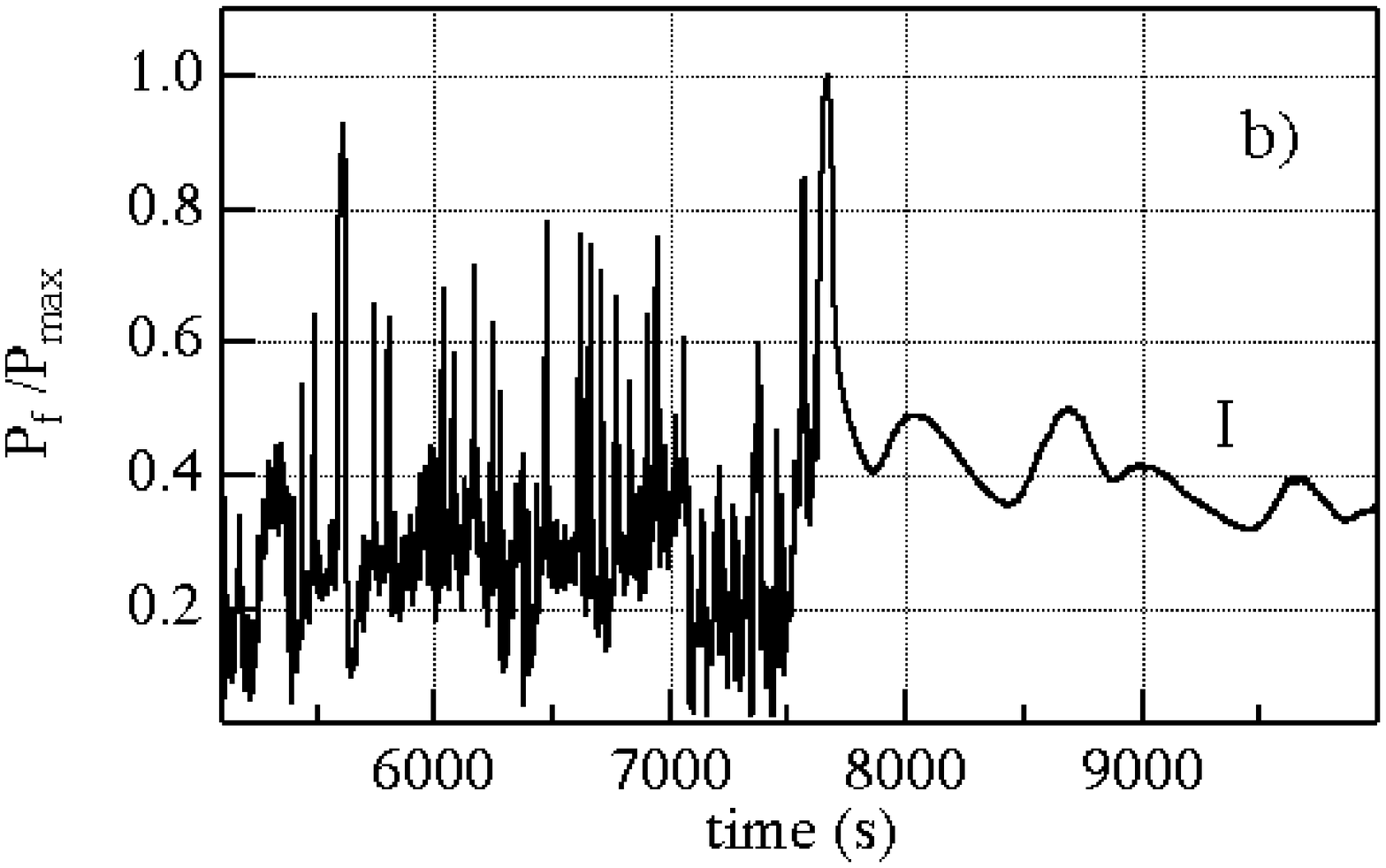}}
\end{figure}
\vspace{-0.4cm}
\parbox[t]{16cm}{\small FIG.4 \ \ The total power dissipated by the 
fluid $P_{f}=IV_{f}$, normalized by the maximum power 
$P_{max}=V_{0}^{2}/4R_{s}$, is plotted vs time. The curves G and I 
are derived from curves B and D (Fig.3) respectively. The curve H is 
calculated from an I(t) curve measured after the curve B.}   
\vspace{0.3cm}

\pagebreak
\begin{figure}[h]
\centerline{ \epsfxsize=8cm \epsfbox{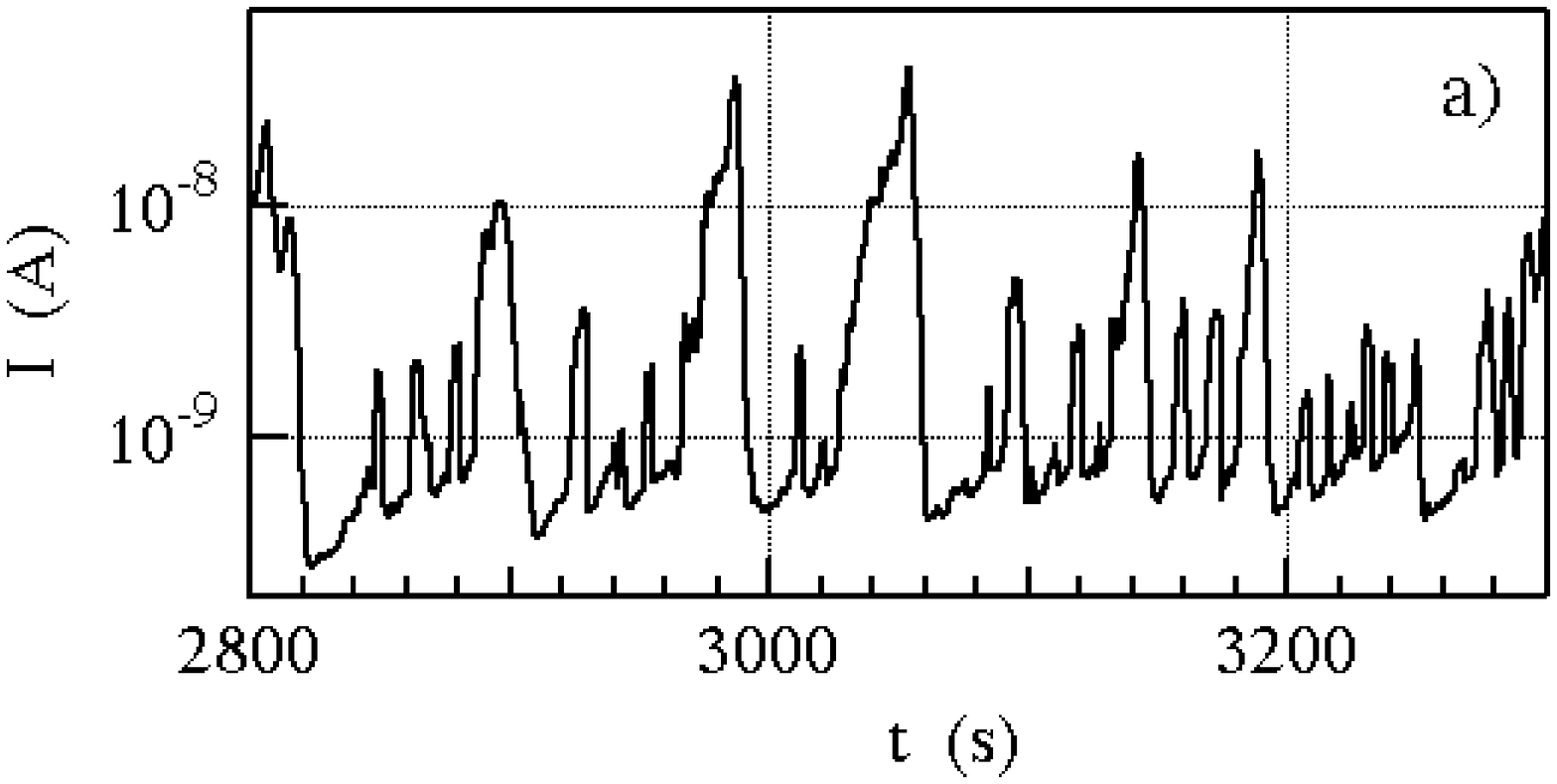}}
\centerline{ \epsfxsize=8cm \epsfbox{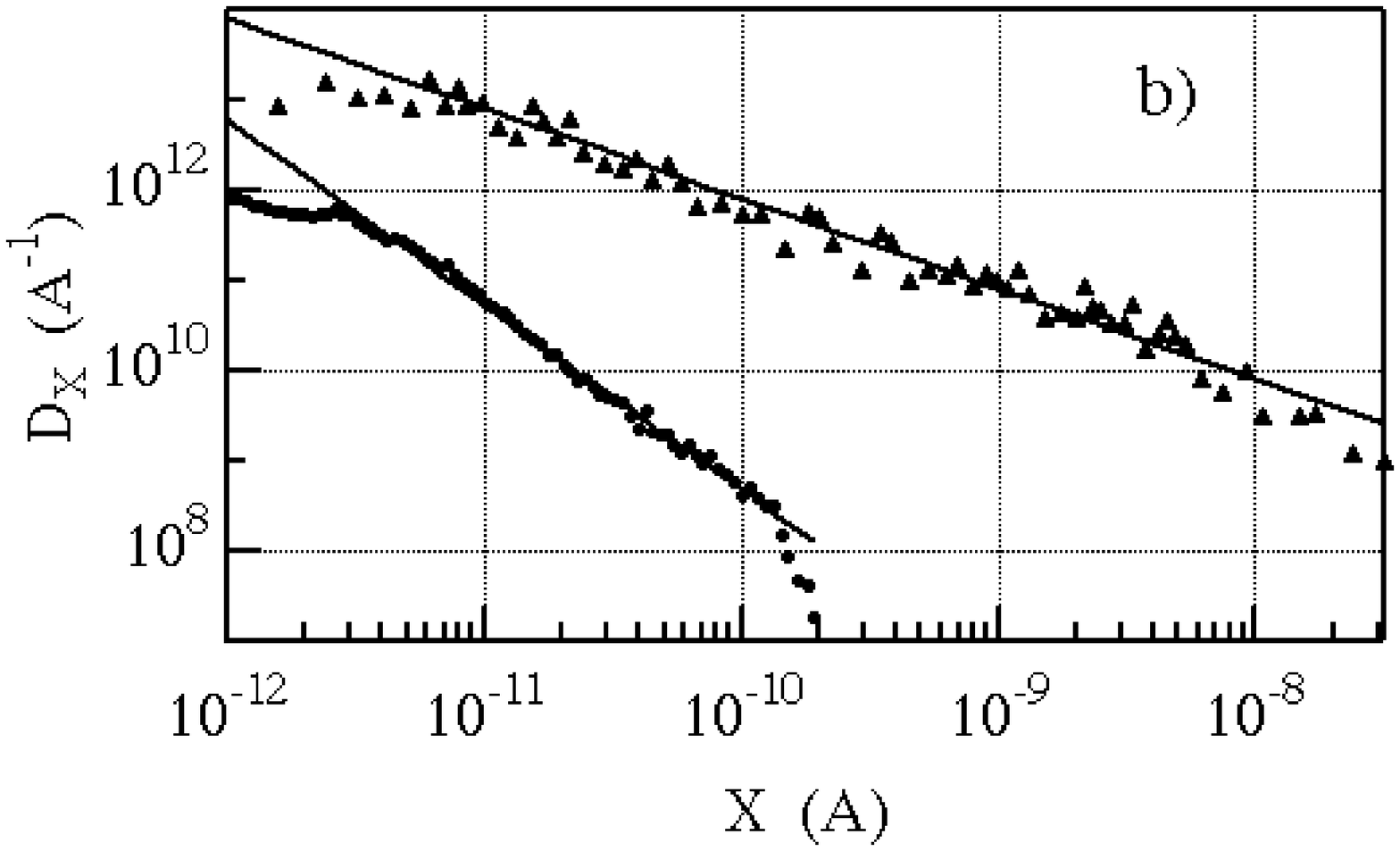}}
\centerline{ \epsfxsize=8cm \epsfbox{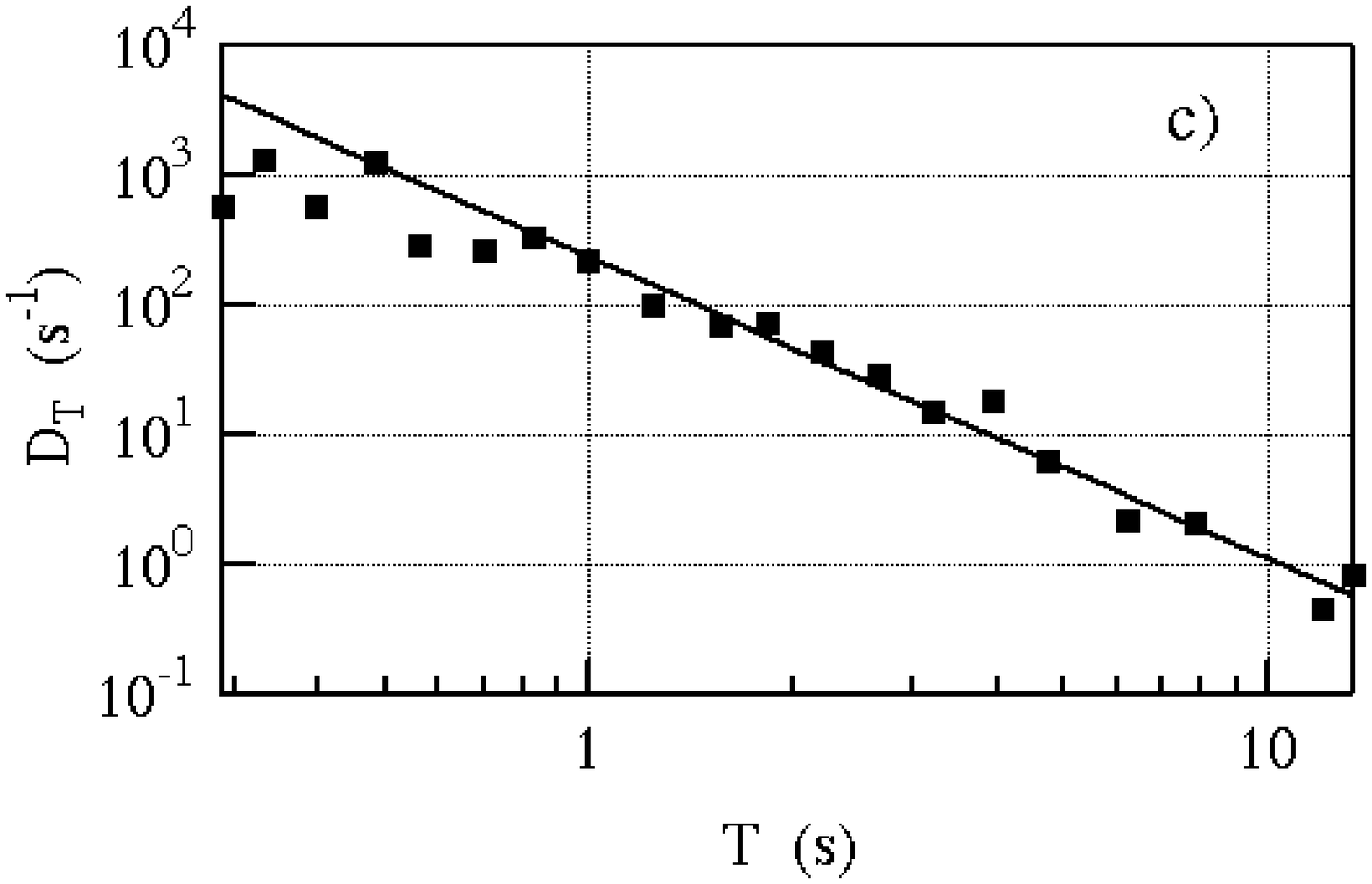}}
\end{figure}
\vspace{-0.4cm}
\parbox[t]{16cm}{\small FIG.5 \ \ (a) A small segment of an evolution curve 
of the type ES3 measured at $V_{0}=100V$ and $R_{s}=2.8 M\Omega$. 
(b) The top curve represents the distribution of avalanche sizes ($X$) 
calculated from the experimental time dependence shown in (a). 
The straight line fit gives the probability density as $D_{X}\sim X^{-\alpha}$ 
with $\alpha = 1$. The bottom curve shows a model-generated 
distribution of avalanche sizes. In this case the exponent in the 
power law is $\alpha\approx 2$. The bottom curve was shifted from its original 
position for clarity. (c) Distribution of avalanche durations ($T$) found 
experimentally. The straight line is $D_{T}\sim T^{-\beta}$. 
The exponent of the power-law fit is $\beta\approx 2.3$.}   
\vspace{0.3cm}

\pagebreak
\begin{figure}[h]
\centerline{ \epsfxsize=8cm \epsfbox{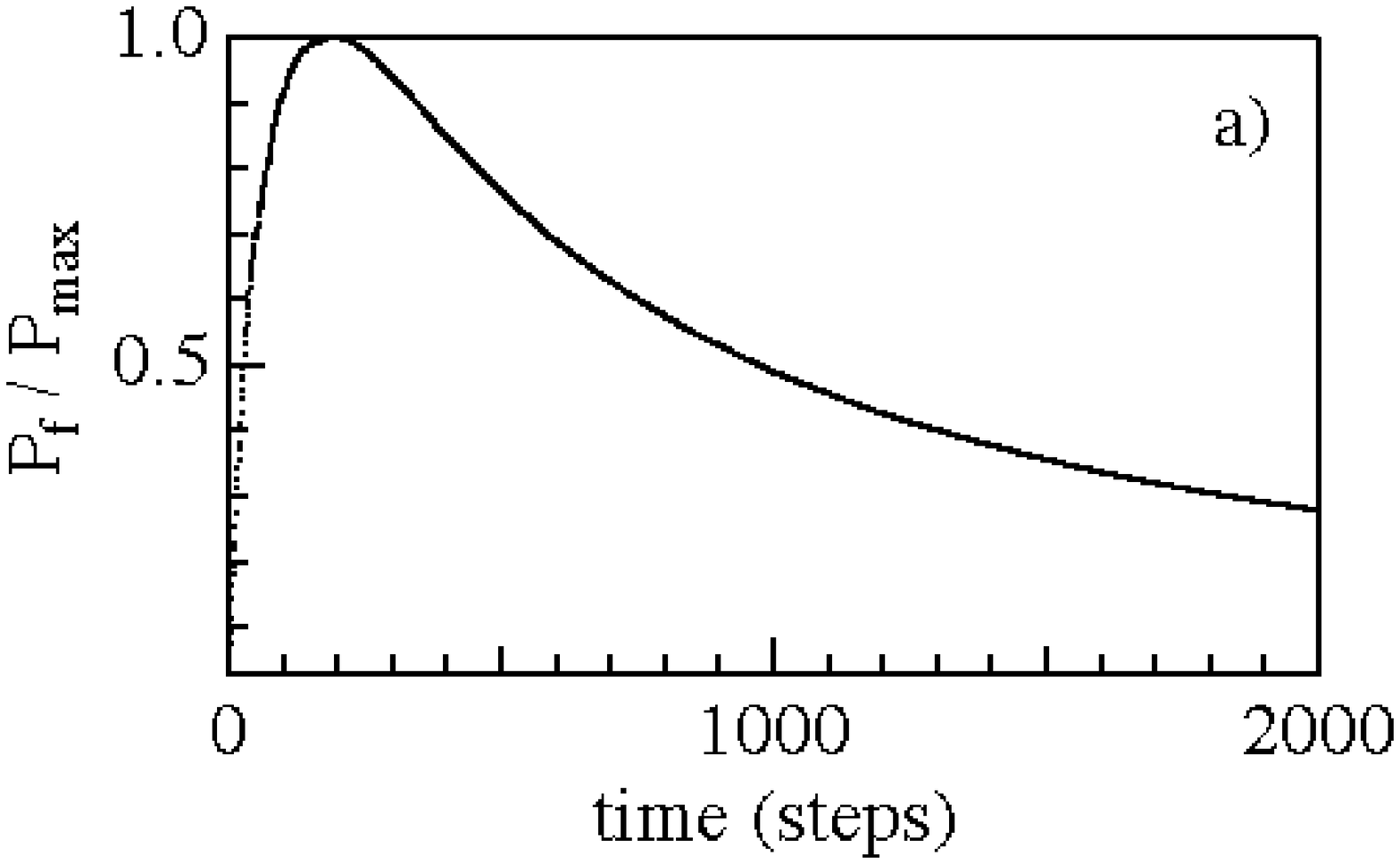}}
\centerline{ \epsfxsize=8cm \epsfbox{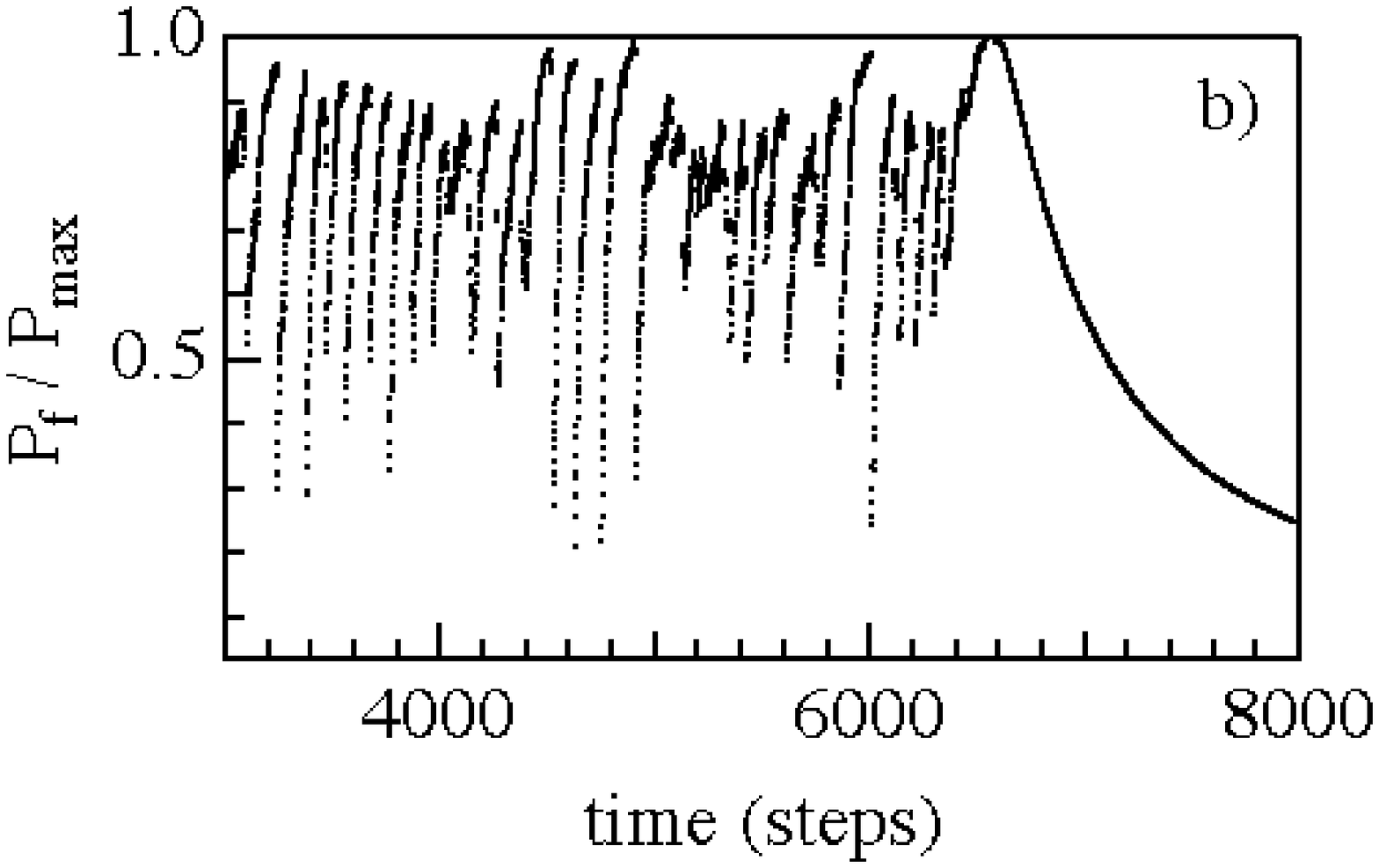}}
\centerline{ \epsfxsize=8cm \epsfbox{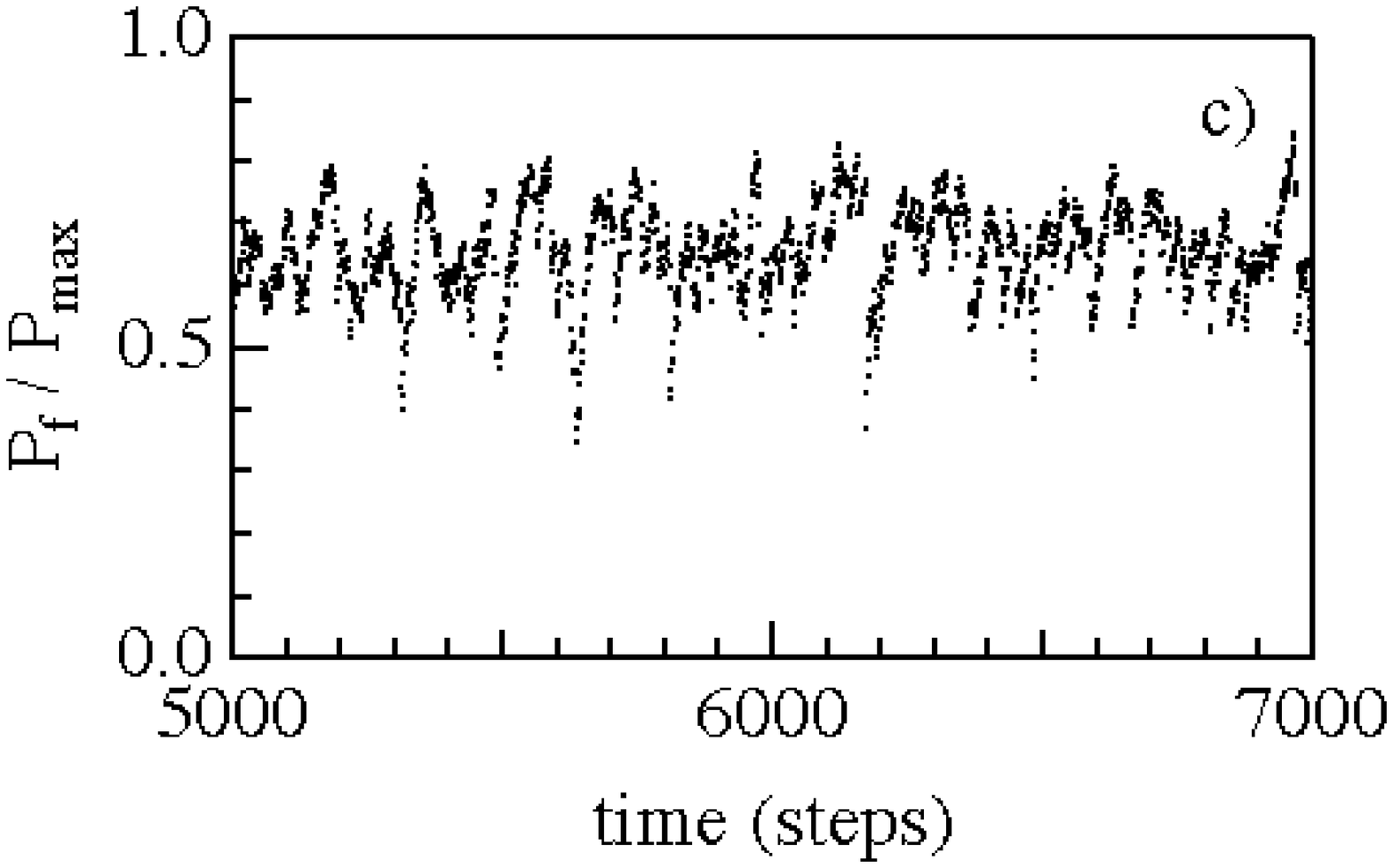}}
\end{figure}
\vspace{-0.4cm}
\parbox[t]{16cm}{FIG.6 \ \ Three examples of model-generated evolution curves. 
All parameters are the same except the growth rate ($STEP$) of the 
dissipative elements. This parameter has the value $STEP=0.2/NR_{s}$, 
$0.3/NR_{S}$, and $0.6/NR_{S}$ for curves a), b), and c), respectively. 
The other parameters are $N=80$, $N_{1}=5$, and $P_{c}=1.8P_{max}/N$.}   
\vspace{0.3cm}

\end{document}